\documentstyle[epsf]{article}

\input epsf
\begin{document}

\vspace*{15mm}

\begin{center}
{\LARGE Test Result of Time-Of-Propagation Cherenkov Counter}~\footnote{Talk given at the 7th  International Conference on Instrumentation for Colliding Beam Physics, INSTR99, November 15-19, 1999, Hamamatsu, Japan.} \\

\vspace*{3mm}

Takayoshi Ohshima \\

\vspace*{3mm}

Physics Department, Nagoya University \\
Chikusa, Furo, Nagoya 464-8602, Japan \\
\end{center}

\vspace*{5mm}

{\small
A new concept concerning Cherenkov detector for particle identification by 
means of measuring both the Time-of-Propagation (TOP) and horizontal emission
angle ($\Phi$) of Cherenkov photons is described here.  Some R\&D works are
also reported. \\}

\section{Introduction}
\par
Measurement of Cherenkov ring image requires two-dimensional photon
information such as x and y coordinates as RICH and DIRC do
\cite{RICH, DIRC}. 
With the use of a quartz bar as a Cherenkov radiator and also a
light-guide like the
DIRC counter \cite{DIRC}, a combination of Time-Of-Propagation (TOP)  of
Cherenkov photons to a bar-end and their emission angles at the bar-end
also provide the ring image information. Here we briefly describe the
principle of such a device, named TOP-counter, ( its detail is
cited in \cite{TOP} ) and explain some results of its R\&D works. The specific
aspect of this counter is its compactness relying upon a horizontal
focussing approach described below.  We intend to develop this counter
in a bid to upgrade the BELLE pid detector.  

\begin{figure}[h]
 \begin{center}
  \leavevmode
       \epsfxsize 10.cm
       \epsfbox{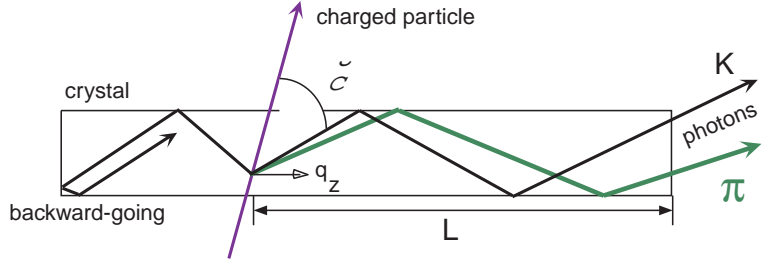} 
\caption{Side view of propagating photons. TOP is inversely proportional
to z-component $q_z$ of the light-velocity: TOP=$\frac{L\times
n(\lambda)}{c q_z}= 4.90(\rm{ns})\times\frac{L(\rm{m})}{q_z}$. $\theta_{\rm{C}}$ is the
Cherenkov angle and $L$ is the particle injection position from the
bar-end. At the opposite-end, a mirror is placed to reflect 
the BW photons. }
\label{fig:Fig.1}
 \end{center}
\end{figure}

Figure~\ref{fig:Fig.1} illustrates a side view of Cherenkov photons propagating a
quartz bar. TOP is inversely proportional to z(quartz-axis
direction)-component of the light-velocity, which produces TOP
differences of, for instance, about 100 ps or more for normal incident 4
GeV/c $K$ and $\pi$ at 2 m long propagation. The TOP difference is a
function of photon's horizontal (x-z plane) emission angle ($\Phi$). 
Time measurement for a single photon with a 100 ps resolution 
provides 1 $\sigma$ separation, and therefore expected number of 30 photons in this case
give us, briefly speaking, a factor of $\sqrt{30}$ times higher separation.
Furthermore, a detection of backward-going (BW) photons reflected at the
other-end as seen in the figure enhances, in principle, the separation by
an another factor of $\sqrt{2}$  for normal incident particle. As is easily
noticed, the TOP measurement inevitably includes also the
Time-Of-Flight (TOF) from an interaction-point to the TOP
counter both of whose difference between $K$ and $\pi$ could have 
the same sign with each other in most of
the cases. Adding the TOF information therefore helps the separation, 
as a result, TOP is hereafter defined as TOP+TOF.

\begin{figure*}[t]
 \begin{center}
    \leavevmode 
       \epsfxsize 13cm
       \epsfbox{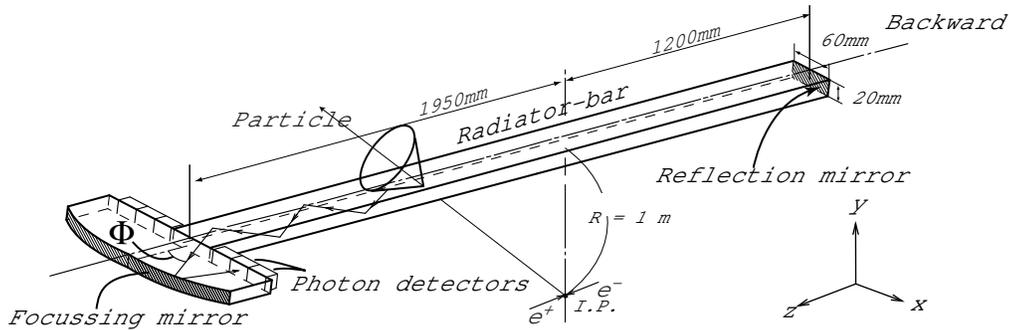} 
 \end{center}
\caption{Structure of the TOP counter. Basic parameters are indicated 
in the figure. 
Since KEKB is an asymmetric collider, the quartz bar is accordingly 
configured z-asymmetric relative to the interaction point (IP). }
\label{fig:Fig.2}
\end{figure*}

In order to estimate the achievable separability of TOP counter, we
optimized its parameters as illustrated in Fig.~\ref{fig:Fig.2}, where the
butterfly-shaped horizontal focussing mirror with an arc radius of 250 mm
was designed to have the $\Phi$-aparture of $\pm$45$^o$ and dispersion of
d$\Phi$/dx=0.5$^o$/1 mm.  Root-mean-square of the focussed accuracy is
$\Delta x \approx\pm$ 0.4 mm. The bar and mirrors are made of synthetic optical
quartz with refractive index (n) of 1.47 at $\lambda$=390 nm.  These counters
are supposed to be placed at 1 m radially away from the interaction point of
KEKB-BELLE to form a cylindrical structure. 

\section{Expected separability}

There are three dominant contributions to TOP measurement: (1) Chromatic
effect of Cherenkov lights, (2) aberration effect of the focussing
mirror and (3) transit time spread (TTS) of photomultiplier tube
(PMT).  
Since BELLE-CDC (Central Drift Chamber) \cite{Haba} provides precise enough track
information such as position, angle and momentum for particles, track
relating ambiguity is about 10 ps which is much smaller than 
the above three
contributions. As a necessary item to be considered, TDC start-signal is
assumed to have a 25 ps uncertainty in the calculation. 

It is worth mentioning that the quartz bar-thickness produces a
harmless effect on the measurement, 
since the variation of a sum of a particle's travel time in a
quartz bar up to Cherenkov radiation point and the photon propagation time to
the bar-end is about 20 ps or less for particles of any incident angle.
Consequently, this contribution is also minute comparing to the others. The
width of the crystal bar, on the other hand, is effectualy nullified, in
principle, due to the horizontal focus, and in practice, within the
achievable focussing accuracy.  This is the reason to choose the horizontal
instead of the vertical focus, otherwise the ring image would grow dim by a
finite size of the bar-width.  Resultantly, both the finite sizes of
thickness and width now can be disregarded, therefore we do not need any
lengthly image projection to nullify the bar cross-section.  

A PMT (Hamamatsu, R5900U-00-L16: linear-arry 16-anodes) is used for R\&D
works without magnetic field (B). Its specific parameters are: Surface area
of 30$\times$30 mm$^{\rm{2}}$, sensitive area of 16$\times$15 mm$^{\rm{2}}$, the
anode size of 0.8 mm-wide with 1.0 mm-pitch and 15 mm-long, quantum
efficiency QE of 20-25\%, gain of 2$\times$10$^{\rm{6}}$, risetime of 0.6 ns, and
TTS of $\sigma$=70-80 ps. Specific modification of L16 and development of a
PMT (R6135MOD-L24: Fine-mesh 24-anodes) operable under a magnetic field are
being proceeded in cooperation with Hamamatsu Co. For the latter PMT, a
position resolution of better than 0.5 mm is achieved under B$>$0.2 TG and
TTS of $\sigma$=130 ps is currently realized under B$<$0.6 TG. 

\begin{figure}[t]
 \begin{center}
    \leavevmode 
       \epsfxsize 8cm
       \epsfbox{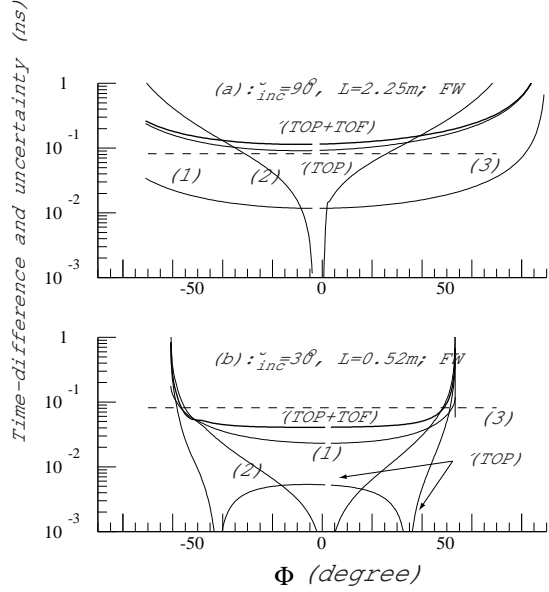} 
\caption{TOP difference and three dominant contributions for 4 GeV/c $K$ and
$\pi$. The counter is supposed to be configured at KEKB-BELLE under
B=1.5 TG, as shown in Fig.~\ref{fig:Fig.2}, and only FW photons 
are detected. 
$\delta$(TOP) and $\delta$(TOP+TOF) are the difference of respective
times between $K$ and $\pi$, and (1), (2) and (3) are the smearing 
contributions described in the text.}
\label{fig:Fig.3}
 \end{center}
\end{figure}

Calculated TOP differences between 4 GeV/c $K$ and $\pi$ and the
above-mentioned three contributions are illustrated for two cases in 
Fig.~\ref{fig:Fig.3},
where TTS is set as 80 ps to include other small
uncertainties such as the start-signal. When the particle incident 
polar angle ($\theta_{\rm{inc}}$) gets around or smaller than 40$^o$, 
TOP difference reverses its sign
against the TOF difference, as seen in Fig.~\ref{fig:Fig.3}(b), 
and the separability power reduces a bit. 
While the expected number of detectable forward-going (FW) 
photons is at an average 35 and 115 at (a) and (b), respectively, 
only the early arrived photons at the individual anodes are taken 
into account for the time measurement. 
When the BW photons are also regarded for detection, 
they come more than 15 ns later than FW photons which are
widely separated enough for measurement to take place and for
distinguishing between each other. 

\begin{figure}[h]
 \begin{center}
    \leavevmode 
       \epsfxsize 8cm
       \epsfbox{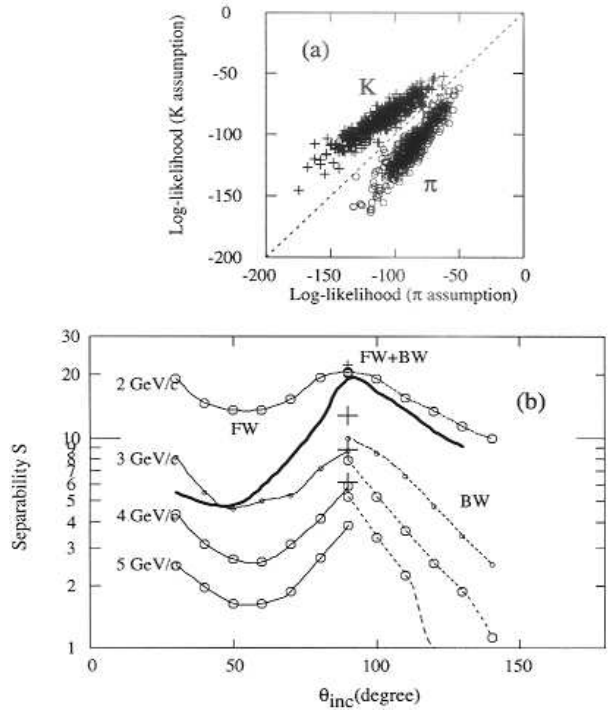} 
\caption{(a) shows a Log-Likelihood distribution with FW photon detection for
4 GeV/c $K$ and $\pi$ with $\theta_{\rm{inc}}$=90$^o$, where the horizontal and
vertical axes correspond to the $\pi$ and $K$ hypotheses to the track,
respectively. Resulted separability is $S$=5.7. (b) is the calculated $S$
in a case of BELLE application, where thin and dotted lines are for only FW
and BW photon detections, respectively, and crosses at
$\theta_{\rm{inc}}$=90$^o$ are by detection of both FW and BW photon. Thick line
is the momentum and polar-angle relation of $\pi$'s in
$B\rightarrow\pi\pi$. }
\label{fig:Fig.4}
 \end{center}
\end{figure}

As a sample of simulation study, Fig.~\ref{fig:Fig.4}(a) shows a Log-Likelihood
distribution in a case of the FW photon detection for 4 GeV/c $K$ and $\pi$
with $\theta_{\rm{inc}}$=90$^o$. 
Resulted separability is $S(=\sqrt{2\Delta ln
\cal{L}})$=5.7. Over-all expected $\pi/ K$ separability is shown in
Fig.~\ref{fig:Fig.4}(b) in the case of BELLE configuration. 
High momentum limit is indicated
by a thick line for the pions in $B\rightarrow\pi\pi$ decay. 
It is found 
that $S>$5 is achieved at any barrel region of $\theta_{\rm{inc}}$=30$^o$-130$^o$.

\section{Beam Test}

A test counter of 1 m long quartz bar was constracted with the structure as
described in Fig.~\ref{fig:Fig.2} but an absorptive filter, 
instead of a reflection mirror, for
BW photons at the bar-end is prepared. Six L16 PMTs (96 anode
channels in total) were attached at the mirror. Since the photoelectron
detection efficiency  of L16 PMT is about 1/2 and an effective mirror surface
coverage by six PMTs with our configuration is approximately 40\%, 
the total photon detection efficiency, besides PMT's QE, is nearly 20\%. 
The above photon insensitive area, most of which is the structual space of PMT, would
reflect the photons and resultantly hit other wrong anodes. To avoid this
phenomena, absorptive filters were inserted in front of such areas.
Measurement was performed using $\pi^-$ beam at KEK-PS. 

\begin{figure}[h]
 \begin{center}
    \leavevmode 
       \epsfxsize 6.5cm
       \epsfbox{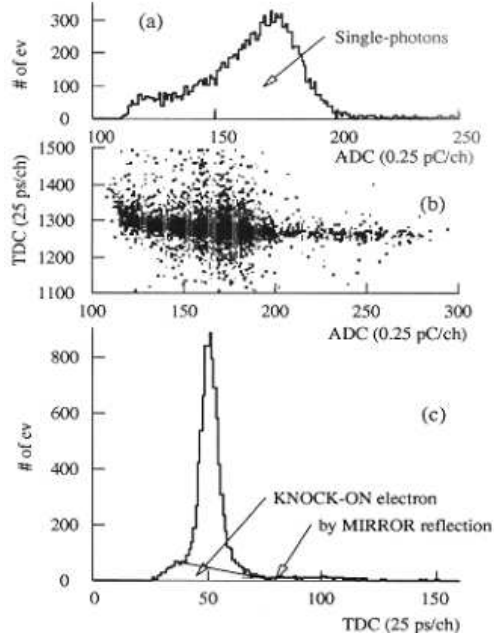} 
\caption{ADC and TDC distributions of 20th channel  for the normal incident 2
GeV/c $\pi$'s with $L$=0.02 m. Timewalk correction was applied in (c) and the
time resolution of trigger uncertainty subtracted is $\sigma$=80 ps. }
\label{fig:Fig.5}
 \end{center}
\end{figure}

\begin{figure}[h]
 \begin{center}
    \leavevmode 
       \epsfxsize 8.0cm
       \epsfbox{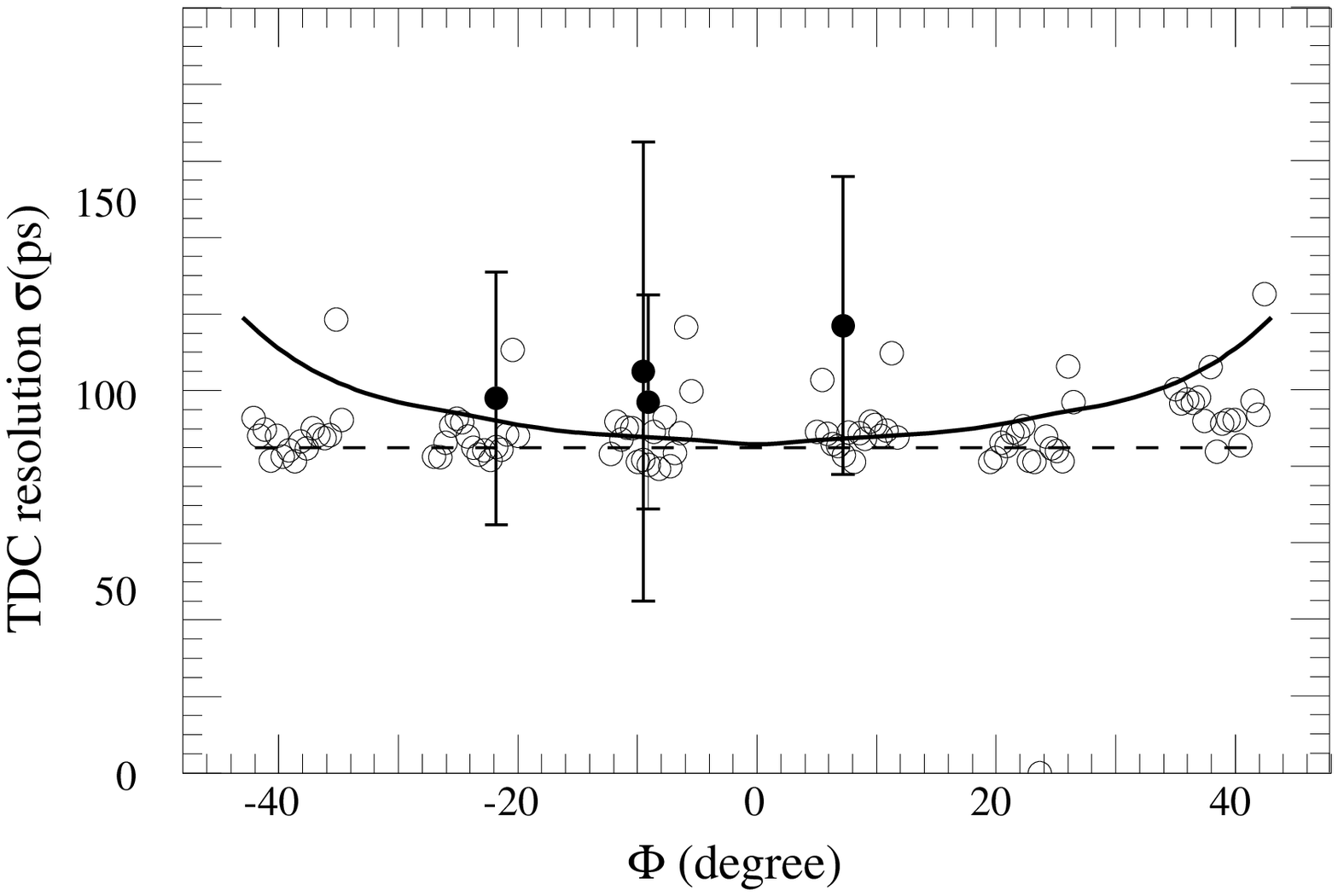} 
\caption{Measured time-resolutions for normal incident 2 GeV/c $\pi$'s.  
Open circles are obtained at $L$=0.02 m. 
The closed circles are obtained at $L$=1 m as described
in the text and the curve is the expected one based on the measurement
of $L$=0.02 m.}
\label{fig:Fig.6}
 \end{center}
\end{figure}

First, beam was tuned to normally hit the counter at $L$=0.02 m.  
Recorded data are shown in Fig.~\ref{fig:Fig.5}. Single
photon peak is clearly seen in ADC spectrum. 
Besides Cherenkov
photons, two small contributions of knock-on electrons and reflected photons
are found on TDC spectrum. Resultant time resolution over all 96 channels
is about $\sigma$=85 ps, as plotted in Fig.~\ref{fig:Fig.6}. Since the chromatic
contribution can be ignored at this configuration, the resulted resolution
is dominated by TTS of L16 PMT. 

\begin{figure}[h]
 \begin{center}
    \leavevmode 
       \epsfysize 11cm
       \epsfbox{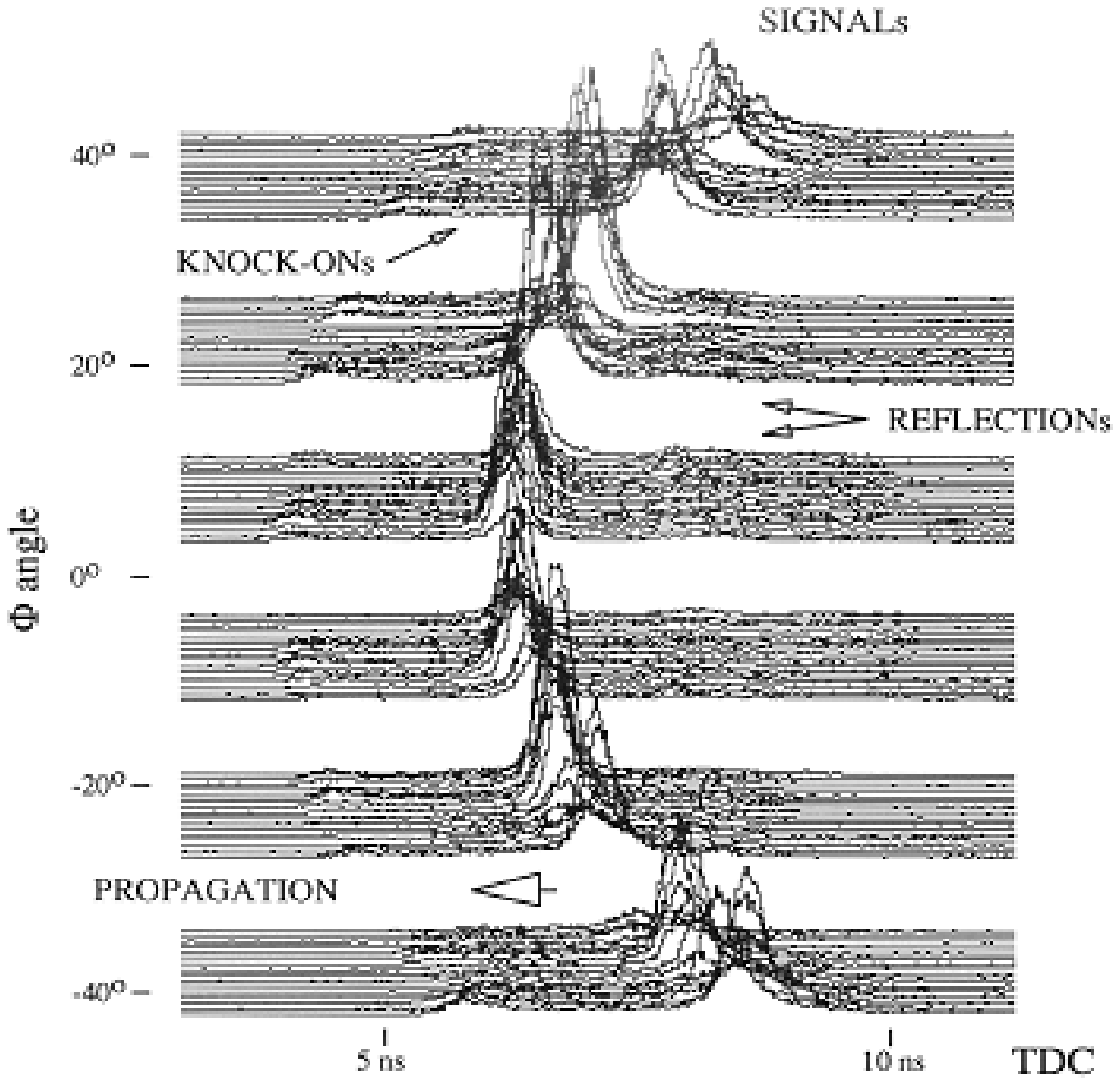} 
\caption{Cherenkov ring image measured by TOP counter for normal incident 4
GeV/c $\pi$'s at $L$=1 m. }
\label{fig:Fig.7}
 \end{center}
\end{figure}

Next, beam position was moved to $L$=1 m and three different momenta of
1.1. 2 and 4 GeV/c were set. Expected number of fired anodes was around 6,
while we observed 6.3 at an average including both the contributions from
the knock-on electron and reflected photons for individual three different
momenta. Cherenkov ring image is clearly observed as a function of
$\Phi$-angle, as seen in Fig.~\ref{fig:Fig.7}. In order to extract the resolution in this
case, a simple tricky analysis had to be applied, because the beam
divergence defined by trigger counters were not sufficiently small enough
as expected at the BELLE detector system to make its contribution
ineffectual.  
That is, the triggered samples are required to have a
signal at a certain channel, for example, 27th channel, within 
the first 150 ps part of the measured raw time distribution of 
350 ps (FWHM). 
This bias would restrict the beam divergence somehow but not
explicitly. Thus obtained resolutions are plotted in Fig.~\ref{fig:Fig.6}: 
Fairly good agreement with the expectation can be seen.  
The parabolic rise of the calculated resolution at 
large $\Phi$ is due to the aberration effect of the mirror 
rather than the chromatic contribution of the Cherenkov light 
at $L$=1 m case.

\section{Summary}

TOP counter is quite compact and has high separability. Due to the
horizontal focussing and thin radiator thickness, the size of quartz bar's
cross-section can be disregarded so that it does not need a large standoff
projection space such as DIRC. It is still at an early R\&D stage and needs
more essential studies as mentioned below. 

First, confirmation of basic TOP behavior, especially the performance at
$L$= 1 m or longer distances, should be done using tracking chambers at next
beam test. 

Increasing the detected number of photons is the most important issue and
two approaches for enlarging the sensitive area are being examined: One way
is to use a light-guide, and the other way is to develop L16 PMT. When a
way to succesfully collect sufficient number of photons is established, TOP
counter can be used as a real detector under certain experimental
condition such as, for instance, fixed target experiment with no magnetic field. It
needs much less space comparing to Gas Cherenkov counter, and can be
configured to make the counter normal to incident particles so that the
separability is enhanced by detecting both the FW and BW photons. 

In order to utilize the TOP counter as the next BELLE pid detector, the
second most important issue is to develop a single photon and position
sensitive, high time-resolving detector operational under a magnetic
field of 1.5 TG. R\&D work of L24 PMT is being earnestly proceeded so that a
successful outcome can be within our grasp in the near future. \\

\par
This work is supported by Grant-in-Aid for Scientific Reasearch on 
Priority Areas (Physics of CP violation) from the Ministry of 
Education, Science, and Culture of Japan.

\end{document}